\documentclass[aps,twocolumn]{revtex4-1}
\usepackage{epsfig}
\usepackage{color}
\usepackage{amsmath}
\usepackage[retainorgcmds]{IEEEtrantools}
\usepackage{amssymb}
\usepackage{graphicx}
\usepackage{wasysym}
\usepackage{esint}
\begin{document}
\title{Entropy production as tool for characterizing nonequilibrium phase transitions}

\author{C. E. Fern\'andez Noa, Pedro E. Harunari, M. J. de Oliveira and 
C. E. Fiore}
\affiliation{ Instituto de F\'{\i}sica da Universidade de S\~{a}o Paulo, 
05314-970 S\~{a}o Paulo,  Brazil }

\date{\today} 
\begin{abstract}
  Nonequilibrium phase transitions can be typified in a
  similar way  to equilibrium
  systems, for instance, by the use of the order parameter.
However, this characterization hides the irreversible character of the
dynamics as well as its influence on the phase transition properties. Entropy
production has revealed to be an important concept for filling this gap since it
vanishes identically for equilibrium systems and is positive for
the nonequilibrium case. Based on distinct and general arguments,
the characterization of phase
transitions in terms of the entropy production is presented.
Analysis  for discontinuous and continuous phase transitions has been
undertaken by taking  regular and complex topologies within the
framework of mean field theory (MFT) and beyond the MFT.
A general  description of entropy production portraits
for  $Z_2$ (``up-down'') symmetry systems under the MFT is presented.
Our main result is that a given  phase transition, whether
continuous or discontinuous has a specific entropy
production hallmark.
 Our predictions are exemplified by an icon system, perhaps the 
simplest nonequilibrium model presenting an
 order-disorder phase transition and spontaneous symmetry breaking: 
the  majority vote model.
 Our work paves the way to a systematic description and 
classification of nonequilibrium phase transitions through  a key 
indicator of system irreversibility.
\end{abstract}

\maketitle

\section{ Introduction}
Thermodynamics states that while  
certain quantities including the  energy   are  ruled by a conservation law,
the  entropy is not  conserved. In the general case
of a system coupled with an environment,
the time variation of entropy $dS/dt$ has two contributions: 
the flux to the reservoir $\Phi$ and  the  entropy production rate $\Pi$
 \cite{prigo,groot}, that is,
\begin{equation}
  \frac{dS}{dt}=\Pi(t)-\Phi(t).
  \end{equation}
Since in the steady state the time variation of $S$ vanishes, $dS/dt=0$, 
 $\Pi=\Phi$ and  all entropy produced must be delivered 
to the environment. 

The entropy production has been the subject of 
considerable interest
in   physics \cite{seinf,tome1,barato,tome2,tome3},  population
dynamics \cite{andrae}, biological systems \cite{mandal}, experimental
verification \cite{landi} and others. 
 A microscopic definition of entropy production,
 in the realm  of systems described by a  master
 equation,  is given by the Schnakenberg expression \cite{schn}: 
\begin{eqnarray}
\Pi(t)=\frac{k_B}{2}\sum_{ij}\{W_{ji}P_i(t)-W_{ij}P_j(t)\}\ln\frac{W_{ji}P_i(t)}{W_{ij}P_j(t)},
\label{schne}
\end{eqnarray}
where  $W_{ji}$ is the transition rate from the state $i$ to state $j$
with associated probability $P_i(t)$ at the time $t$ and  $W_{ij}$ denotes
the reverse transition rate.
Eq. (\ref{schne}) implies that $\Pi(t)$
 is always  non negative because $(x-y)\ln (x/y) \ge 0$,  vanishing  when  the detailed balance
 $W_{ij}P_j-W_{ji}P_i=0$ is fulfilled. Thus it distinguishes
 equilibrium  from  nonequilibrium systems.
 Defining the nonequilibrium entropy by
 $S(t)=-k_B\sum_{i}P_i(t)\ln P_i(t)$, a microscopic 
 relation for the flux $\Phi(t)$ is obtained:
\begin{equation}
  \label{prod}
	 \Phi(t)=k_B\sum_{i,j}W_{ij}\ln \frac{W_{ij}}{W_{ji}}P_{j}(t).
\end{equation}

Eq. (\ref{prod}) constitutes an alternative (and advantageous) formula
for evaluating the
steady entropy production, since it corresponds to an average
that can be evaluated  from the 
transition rates and  it will be subject of   analysis in the present paper. 

Despite the recent  advances of stochastic thermodynamics, a fundamental question  is whether  
entropy production can be utilized as a  reliable tool for 
typifying nonequilibrium  phase transitions.
Different   studies have been undertaken 
 in this direction \cite{tome1,barato,wetting,andrae,gaspard,qian,imparato,shim,esposito1,esposito2}.
  Some of them \cite{tome1,barato,wetting,andrae}
 indicate that   continuous phase transitions
can be identified  by a divergence of the first 
derivative of $\Pi$ whose associated exponent plays an analogous
role to the specific heat.
Other features, such as stochastic thermodynamics of many particle
systems at phase transitions to a  synchronized regime have also
been investigated \cite{imparato,esposito1,esposito2}.
Despite such a progress,   a theoretical description of the   
entropy production at  phase transition regimes, mainly
in the context of discontinuous phase transition, has  not been satisfactorily
established yet.

In this paper we present a characterization of phase
transitions in terms of the entropy production. 
Our study embraces the analysis
of continuous and discontinuous phase transitions within the
 framework of mean field theory (MFT) and beyond  MFT.
It  is based on general considerations
about the probability distribution related to the
phase coexistence. The description of continuous 
phase transition takes into account the extension of  finite-size scaling ideas
and hyperscaling relations to nonequilibrium systems. 
A
general  description of entropy production
for  $Z_2$ (``up-down'') symmetry systems in the realm of MFT is presented.
Our main result is that a given  phase transition, whether
continuous or discontinuous has a specific entropy
production signature.
 As an example of our theoretical prescriptions, we shall
 consider the majority vote (MV) model with inertia \cite{mario92,chen1,chen2,pedro,jesus}. It constitutes an ideal laboratory, since it presents continuous and discontinuous phase transition in both regular \cite{mario92,jesus} and complex
 structures \cite{chen1,chen2,pedro} displaying quite distinct features and
  universality classes. Thus, the existence of different
 entropy production hallmarks at phase transition regimes can be conveniently
 compared with those obtained from  order parameter analysis.

 This paper is organized as follows: In Sec. II  we derive a general mean-field description for  $Z_2$ (``up-down'')
 symmetry systems. Sec. III presents a description of entropy production
 at phase transition regimes beyond the MFT. In Sec. IV, we exemplify
 our theoretical findings in the inertial MV model and Conclusions are performed
 in Sec. V.

\section{General mean-field description for  $Z_2$ (``up-down'')
symmetry systems}
We are dealing with phase transitions in systems with up-down symmetry.
Heuristically, a continuous phase transition in such class of models
is described
by the general logistic  order-parameter equation:
\begin{equation}
  \frac{d}{dt}m=a(q-q_c)m-bm^3,
  \label{op1a}
\end{equation}
where $q$ denotes the control parameter and  $a$ and $b$
are positive constants. It has two steady solutions: 
$m^{(D)}=0$ (disordered phase)  and
$m^{(S)}=\pm \sqrt{a(q-q_c)/b}$ (ordered phase),
stable for low and large values of $q$, respectively.
The phase transition follows the mean-field exponent $\beta_{mf}=1/2$
and $m$ vanishes as $m\sim e^{a(q-q_c)t}$ for $q<q_c$ when
$m\ll 1$.
Conversely,  one requires the inclusion of an additional term $+cm^5$
for reporting  discontinuous
phase transitions, leading to the following expression
\cite{jesus}:
\begin{equation}
  \frac{d}{dt}m=a(q_b-q)m-bm^3+cm^5,
  \label{op1b}
\end{equation} 
where $c>0$ \cite{jesus}.   It exhibits three steady state
solutions $m$: 
$m^{(D)}=0,\mbox{ }m^{(S)}\mbox{ and }m^{(U)}.$
At $q=q_f=(b^2/4ac)-q_b$,
$m$ jumps from $m_1\equiv m^{(S)}(q_{f})$ to $m^{(D)}=0$. For $q>q_f$,
 $m$ behaves  as 
 $m \sim e^{a(q_b-q)t}$ for $m_0\ll1$ irrespective the initial condition
 $m_0>0$. 
The frontier $q=q_b $ separates  the exponential
vanishing of $m \sim e^{a(q_b-q)t}$ ($q>q_b$) from the convergence to a well 
definite $m_2\equiv m^{(S)}(q)$ ($q<q_b$) when $m_0\ll 1$.
For
$q_b<q<q_f$ (hysteretic branch), 
$m$ behaves  as follows:
 $m(t\rightarrow \infty)\rightarrow m^{(D)}$
if $m_0<m^{(U)}$, $m(t\rightarrow \infty)\rightarrow m^{(S)}$
if $m_0>m^{(U)}$ and only for $m_0=m^{(U)}$ one has
$m(t\rightarrow \infty)\rightarrow m^{(U)}$. For this reason
$m^{(U)}$ is an unstable solution.

Since the above phenomenological relations hide the irreversible
character which we are interested,  we derive a general expression for
the entropy production
taking into account 
a generic dynamics with up-down symmetry.
Each site $i$  of an arbitrary lattice topology is attached
to a spin variable $\sigma_i$ that assumes the values 
$\pm 1$. The transition rate
is  given by the expression
$w(\sigma_i)=\frac{1}{2}[1-q\sigma_ig(X)]$, with $q$ denoting the
control parameter and $g(X)$ expressing
the generic dependence on a local neighborhood of $k$  spins.
Only  two assumptions regarding $g(X)$ are required.
The first is that due to the $Z_2$
symmetry, it depends 
on the sign of the local spin neighborhood (odd function).
Also, taking into account that $w(\sigma_i)$ is constrained between $0$
and $1$,  the product $|q g(X)| \le 1$
for all values of $X$. These  assumptions allow us to rewrite
$g(X)$  as $g(X)=|g(X)|S(X)$, where  $S(X)$ denotes the sign function:
${\rm sign}(X)=1(-1)$ and $0$, according to $X>0(<0)$
and $X=0$, respectively, where $|g(X)|$ gets restricted
between $0$ and $|g(k)|$.

From the master equation, one finds that
the time evolution of  order parameter 
$m=\langle \sigma_i\rangle$  is given by
\begin{equation}
  \frac{d}{dt}\langle \sigma_i\rangle=-2\langle \sigma_iw(\sigma_i)\rangle.
  \label{eap}
\end{equation}
In the steady state  $m=q\langle|g(X)|S(X) \rangle$.
For the evaluation  of $\Pi$, one requires the calculation of
$w_i (\sigma)\ln[w_i (\sigma)/w_i (\sigma^j)]$
given by
\begin{equation}
\frac{1}{2}\left[\sigma_i S(X)-q|g(X)|S^2(X)\right]\ln\frac{1-q|g(X)|}{1+q|g(X)|}.
\label{ep2}
\end{equation}
The reverse transition rate $w_i (\sigma^j)$ was obtained
by performing the transformation $\sigma_i \rightarrow -\sigma_i$ resulting
in $w_i (\sigma^j)=\frac{1}{2}[1+q\sigma_ig(X)]$.
 The one-site MFT consists of rewriting the
joint probability $P(\sigma_i,...,\sigma_{k})$
as a product of one-site probabilities $P(\sigma_i)...P(\sigma_{k})$, 
from which one derives closed relations for  the correlations
from which we obtain the
properties as function of the control parameters.
Since the main marks of critical and discontinuous
phase transitions are not expected to depend 
on the particularities of $g(X)$,  it is reasonable,
within the MFT,
 to replace the averages  in terms of an effective ${\bar g}$
 given by 
\begin{equation}
m=q\langle |g(X)|S(X)\rangle \rightarrow q{\bar g}\langle S(X)\rangle,
\end{equation}
 \begin{equation}
 \frac{1}{2}\left\langle\sigma_i S(X)\ln\frac{1-q|g(X)|}{1+q|g(X)|}\right\rangle\rightarrow\frac{1}{2}\ln\frac{1-q{\bar g}}{1+q{\bar g}}\langle\sigma_i S(X)\rangle,
\end{equation}
 and
  \begin{equation}
 \frac{1}{2}\left\langle|g(X)|S^2(X)\ln\frac{1-q|g(X)|}{1+q|g(X)|}\right\rangle\rightarrow \frac{{\bar g}}{2}\ln\frac{1-q{\bar g}}{1+q{\bar g}}\langle S^2(X)\rangle.
\end{equation}
At this level of approximation the  steady entropy production then reads
\begin{equation}
\Pi=
\frac{1}{2}\ln\frac{1-q{\bar g}}{1+q{\bar g}}\left[m \langle S(X)\rangle-q{\bar g}\langle S^2(X)\rangle\right].
\label{ep3}
\end{equation}
Above averages are calculated by decomposing the mean sign function   
in two parts:
\begin{equation}
  \langle S(X)\rangle=\langle S(X_+)\rangle-\langle S(X_-)\rangle,
  \label{sig1}
  \end{equation}
 and
\begin{equation}
  \langle S^{2}(X)\rangle=\langle S(X_+)\rangle+\langle S(X_-)\rangle,
  \label{sig11}
  \end{equation}
 with each term  being approximated by
\begin{equation}
  \langle S(X_\pm)\rangle = \pm\sum_{n=\lceil k/2\rceil}^{k}C_{n}^{k}p_{\pm }^{n}p_{\mp }^{k-n},
  \label{eq4}
\end{equation}
 where $\lceil ...\rceil$ is the ceiling function  
and for $S(X_+)[S(X_-)]$ the term 
$C_{n}^{k}$ takes into account the number of
possibilities of a neighborhood with $n$  spins
in the $+1[-1]$  states 
with associated probabilities $p_{\pm}=(1\pm m)/2$.
Eqs. (\ref{sig1}) and (\ref{sig11}) 
become simpler in the regime of large connectivities.
To see this, we first note that each term of the binomial distribution
  approaches  a Gaussian
  with mean $kp_{\pm }$ and  variance $\sigma^{2}=kp_{+}p_{-}$, so that
 \begin{eqnarray}&&\sum_{n=\lceil k/2 \rceil}^{k}C_{n}^{k}p_{\pm}^{n}p_{\mp}^{k-n} \rightarrow  \frac{1}{\sigma\sqrt{2\pi}}\int_{k/2}^k  e^{-\frac{(\ell-kp_{\pm})^2}{2\sigma^2}}d\ell=\nonumber\\&=&
\frac{1}{2}\sqrt{\pi}\left\{{\rm erf}\left[\frac{k(1-p_{\pm})}{\sqrt{2}\sigma}\right]-{\rm erf}\left[\frac{k(1/2-p_{\pm})}{\sqrt{2}\sigma}\right]\right\},
\end{eqnarray} 
where ${\rm erf(x)}=2\int_{0}^{x}e^{-t^2}dt/\sqrt{\pi}$ denotes the error function.
Since for large $k$, ${\rm erf}[k(1-p_{\pm})/\sqrt{2}\sigma] \rightarrow 1$  ($\langle S^{2}(X)\rangle \rightarrow 1$),
the expressions for $m$ and $\Pi$ read
\begin{equation}
  m=q{\bar g}\left[\mathrm{erf}\Big(\sqrt{\frac{k}{2}}m\Big)\right],
\end{equation}
and 
\begin{equation}
\Pi=
\frac{1}{2}\ln\frac{1-q{\bar g}}{1+q{\bar g}}\left[\frac{m^2}{q {\bar g}}-q{\bar g}\right],
\label{ep3}
\end{equation}
respectively.
At the vicinity
of the critical point $m$ behaves as
$m \sim (q-q_c)^{1/2}$. So that, one reaches the following
expressions for the entropy production: 
\begin{equation}
\Pi \sim 
\frac{1}{2}\ln\frac{1+q{\bar g}}{1-q{\bar g}}\left[\frac{q_c-q}{q {\bar g}}+q{\bar g}\right],
  \label{epc1}
  \end{equation}
for $q<q_c$, and
\begin{equation}
\Pi =
\frac{q{\bar g}}{2}\ln\frac{1+q{\bar g}}{1-q{\bar g}},
  \label{epc2}
  \end{equation}
for $q>q_c$. Hence the entropy production is continuous at the critical
point $q_c$, with
$\Pi_c=\frac{q_c{\bar g}}{2}\ln\frac{1+q_c{\bar g}}{1-q_c{\bar g}}$.
However, its first derivative is discontinuous,
jumping from
\begin{equation}
  \Pi'=\frac{q_c{\bar g}^2}{1-q_c^2{\bar g}^2}+\frac{1-q_c{\bar g}^2}{2q_c{\bar g}}\ln\frac{1-q_c{\bar g}}{1+q_c{\bar g}},
\end{equation}
when $q \rightarrow q_c^{-}$, to
\begin{equation}
  \Pi'=\frac{q_c{\bar g}^2}{1-q_c^2{\bar g}^2}-\frac{{\bar g}}{2}\ln\frac{1-q_c{\bar g}}{1+q_c{\bar g}},
\end{equation}
when $q \rightarrow q_c^{+}$, whose discontinuity
of $-\frac{1}{2q_c{\bar g}}\ln\frac{1-q_c{\bar g}}{1+q_c{\bar g}}$
is associated with the critical exponent $\alpha_{mf}=0$.
Remarkably, having the classical exponents $\beta_{mf}$ and $\gamma_{mf}$
(evaluated from the order-parameter variance \cite{mariobook}), we see that
the hyperscaling relation $\alpha_{mf}+2\beta_{mf}+\gamma_{mf}=2$
is  satisfied,   reinforcing  that the  criticality is signed by  the
jump in the first derivative of $\Pi$, 
in close similarity to the specific heat discontinuity for equilibrium systems.

Above MFT entropy production also predicts correctly the
signatures at discontinuous phase transitions. According to  Eq. (\ref{op1b}),
 $m$ jumps from $m_1\equiv m^{(S)}(q_{f})$ to $0$  at 
$q=q_f=(b^2/4ac)-q_b$ and  thereby from Eq. (\ref{ep3})  
the entropy production will jump from
 \begin{equation}
   \frac{1}{2}\left(q_f{\bar g}-\frac{m_1^2}{q_f{\bar g}}\right)\ln\left[\frac{1+q_f{\bar g}}{1-q_f{\bar g}}\right],
\end{equation}
   to
\begin{equation}
   \frac{q_f{\bar g}}{2}\ln\left[\frac{1+q_f{\bar g}}{1-q_f{\bar g}}\right].
\end{equation}
Conversely  $m$ jumps from $0$ to $m_2\equiv m^{(S)}(q_{b})$ at $q=q_b$
and hence
$\Pi$ will jump from
\begin{equation}
   \frac{q_b{\bar g}}{2}\ln\left[\frac{1+q_b{\bar g}}{1-q_b{\bar g}}\right],
\end{equation}
to
\begin{equation}
   \frac{1}{2}\left(q_b{\bar g}-\frac{m_2^2}{q_b{\bar g}}\right)\ln\left[\frac{1+q_b{\bar g}}{1-q_b{\bar g}}\right].
\end{equation}
The bistable behavior in the entropy production not only
discerns continuous and discontinuous phase transitions  but also it 
properly locates the hysteretic loop.
In the Sec. IV, we show explicit results by taking an example of
system with $Z_2$ symmetry.
\section{Beyond the mean-field theory}
The analysis will be splitted  in three parts:
 discontinuous transitions in  
regular lattices, complex networks and   continuous phase transitions.
\subsection{Discontinuous phase transitions}
\subsubsection{Regular Lattices}
Distinct works  \cite{jesus,pedro,fsize2,chen2} have attested that 
discontinuous phase transitions yield stark differences   in regular and complex networks. In 
the former case, it  emerges through sudden changes of $|m|$,
its variance
$\chi=N[\langle m^{2}\rangle-|m|^{2}]$ and other quantities
whose scaling behavior goes with the system 
volume $N$  (see e.g. 
panels $(b)$-$(d)$ in Fig. \ref{fig2-1}) \cite{jesus,pedro,fsize2}.
 At the vicinity of
 an arbitrary discontinuous phase transition point $q_0$,
in which the correlation length is finite,  the probability 
distribution  can be approximately written down as a sum of two independent 
Gaussians, from which one extracts a scaling
behavior with the system volume  \cite{jesus,fsize,fsize2,challa}.
More specifically, the probability distribution is  given by
$P_{N}(m) = P_{N}^{(o)}(m) + P_{N}^{(d)}(m)$, where 
$P_{N}^{(\alpha)}(m)$ is associated to the phase $\alpha$
(with order-parameter $m_\alpha$):
\begin{equation}
  P_{N}^{(\alpha)}(m) = \frac{\sqrt{N}}{\sqrt{2 \pi}} \,\frac{\exp[N \{\Delta q m -  (m-m_\alpha)^2/(2 \chi_\alpha)\}]} {[F'_o(\Delta q;N) + F'_d(\Delta q;N)]}.
\end{equation}
Parameters  $\chi_\alpha$ and $\Delta q\equiv q_N-q_{0}$ correspond
to the  distribution width  and
the ``distance'' to the coexistence point $q_0$,
 respectively. 
 Although in principle the assumption of two independent Gaussians can not
 describe properly   a ``weak''  discontinuous phase transition, in
 which an overlap between $P_{N}^{(o)}(m)$ and $P_{N}^{(d)}(m)$ is expected,
 its reliability has been verified in several   examples
 of nonequilibrium phase transitions with distinct properties \cite{fsize,fsize2}, even in some cases in which the overlap 
 is observed.

 Despite the steady entropy production   displaying a non-trivial dependence on  the system features and on generic correlations
of type $\langle\sigma_i\rangle$, $\langle\sigma_i\sigma_{i+1}\rangle$, 
$\langle\sigma_i\sigma_{i+1}\sigma_{i+2}\rangle$ and so on, Eq. (\ref{prod})
depicts it as the ensemble average of a fluctuating
quantity, enabling  resorting 
to the central limit theorem ideas. 
The generality of order-parameter  distribution for tackling
the phase coexistence \cite{fsize2} and
Eq. (\ref{prod}) setting up $\Phi$ as
an ensemble average suggests the  extension of a similar relationship
for the  steady entropy production.
More concretely, we assume that
$P_{N}(\phi) = P_{N}^{(o)}(\phi) + P_{N}^{(d)}(\phi)$, where $P_{N}^{(\alpha)}(\phi)$
is given by
\[P_{N}^{(\alpha)}(\phi) = \frac{\sqrt{N}}{\sqrt{2 \pi}} \,\frac{\exp[N \{\Delta q \phi -  (\phi-\phi_\alpha)^2/(2 {\bar \chi_\alpha})\}]} {[F_o(\Delta q;N) + F_d(\Delta q;N)]},\]
where each Gaussian is centered at $\phi_\alpha$ with  ${\bar\chi_\alpha}$
being  the    width of the $\alpha-$th peak.
Given that  $P_{N}(\phi)$ is normalized, each term $F_{o(d)}$ then reads $F_{o(d)}(\Delta q;N) = \sqrt{{\bar \chi_{o(d)}}} \,\exp\left\{N \Delta q\left[\phi_{o(d)} +  {\bar \chi_{o(d)}} \Delta q/2\right]\right\}$.  The steady
entropy production $\Pi=\Phi$ is straightforwardly calculated from
$P_{N}(\phi)$, $\Phi=\int_{-\infty}^{\infty}  \phi P_{N}(\phi)  d\phi$, reading
\begin{equation}
\Pi  = \sum_{\sigma = o,d} \, \frac{(\phi_\sigma + {\bar \chi_\sigma} \Delta q)\,F_\sigma(\Delta q;N)}{F_o(\Delta q;N) +  F_d(\Delta q;N)}.
\label{eqppc}
\end{equation}
Close to the phase  coexistence, in which $\Delta q$ is expected
 to be small, the terms $O(\Delta q)$ dominate
over  $O(\Delta q)^{2}$ and   Eq. (\ref{eqppc})  can be approximately rewritten   as 
\begin{equation}
 \Pi =\frac{\sqrt{{\bar \chi_o}}\phi_o+\sqrt{{\bar \chi_d}}\phi_de^{-N[(\phi_o-\phi_d)\Delta q]}}{\sqrt{{\bar \chi_o}}+\sqrt{{\bar \chi_d}}e^{-N[(\phi_o-\phi_d)\Delta q]}}.
\label{opr}
\end{equation}
Note that the  Eq. (\ref{opr}) reproduces the jump from $\phi_o$($\phi_d$) 
 when $\Delta q\rightarrow 0_{-(+)}$  and $N\rightarrow \infty$ 
(a third reason for assuming  $P_{N}(\phi)$ as a sum of independent Gaussians). Remarkably,  
the curves for different values of $N$ cross at  the transition point $\Delta q=0$
with
\begin{equation}
  \Pi^*=\frac{\sqrt{{\bar \chi_o}}\phi_o+\sqrt{{\bar\chi_d}}\phi_d}{\sqrt{{\bar \chi_o}}+\sqrt{{\bar \chi_d}}}.
  \label{crossing}
\end{equation}
The crossing point  clearly discerns continuous and discontinuous
phase transitions and can be used as an indicator of the phase coexistence,
as shown in Figs. \ref{fig2-1},  \ref{fig5c}  and
in Ref. \cite{stama} (Figs. 7 and 8) for a chemical reaction model.

\subsubsection{Complex networks}
Distinct works \cite{chen2,pedro,prl2011,jesus} have stated that in contrast
 to regular structures,  the phase coexistence  in complex networks is akin
to the MFT  (see e.g Fig. \ref{fig2m}),
whose behavior is generically characterized by  the existence
of a hysteretic loop and bistability.
The order parameter will present a spinodal line in which
along the hysteretic loop the system will converge to one of the possible
steady states depending on the initial configuration.
For locating the ``forward transition'' point $q_f$,
the system is initially placed in an ordered configuration and the tuning
parameter $q$ is increased by an amount $\delta $,
whose final state at $q$ is
used as the initial condition at $q+\delta $ until
the order-parameter discontinuity is viewed.
Conversely, the
``backward transition'' point $q_b$ is pinpointed 
by starting from  the disordered phase and 
decreasing $q$ (also by the increment $\delta $)
until the order-parameter jump takes place.  Entropy production also captures these features, which can be viewed through a general argument for order-disorder phase transitions.
The order-parameter  behaves as
  $\langle\sigma_i\rangle \sim N^{-1/2}$ in the disordered phase and then a 
$n-$th correlation  will behave as
$\langle\sigma_i\sigma_{i+1}...\sigma_{i+n}\rangle \approx \langle\sigma_i\rangle \langle\sigma_{i+1}\rangle...\langle \sigma_{i+n}\rangle=N^{-n/2}$. Hence
in the thermodynamic limit,  all correlations will vanish  
 in the disordered phase and
$\Pi$ will depend solely on control parameters.
Contrariwise,
$\langle\sigma_i\sigma_{i+1}...\sigma_{i+n}\rangle$ presents a 
 well defined (nonzero) value in the ordered phase and
$\Pi$  depends not only on the control parameters but also on correlations.
 So that, the   jumps at $q_f$ (from $m_1\equiv m(q_f)\neq 0$ to 0) and $q_b$ (from $0$
to $m_2\equiv m(q_b)\neq 0$), commonly viewed in terms of order-parameter,
will also be present in the entropy production. The presence of
bistability  implies
that  $ \Phi(t)$  will converge
to one of the two well defined values, since
along the hysteretic branch the system behaves just like  the
disordered or the ordered phase, depending on the initial condition.
Although the above argument is valid for a generic order-disorder
phase transition, it is expected to describe 
phase transitions different  from the order-disorder ones,
provided the order-parameter and 
correlations also present a hysteretic behavior.
Thereby, both cases reveal that the entropy production behavior also embraces
phase coexistence traits commonly treated in terms of the order-parameter.

\subsection{Continuous phase transitions}
Albeit  characterized by the vanishing of the order-parameter $|m|$
 and algebraic divergences of other quantities
at the criticality, the behavior of  quantities  become rounded
due to finite size effects.  According to the standard
finite-size scaling (FSS), they behave as
$|m|=N^{-\beta/\nu}{\tilde f}(N^{1/\nu}|\epsilon|)$,
$\chi=N^{\gamma/\nu}{\tilde g}(N^{1/\nu}|\epsilon|)$ with
${\tilde f}$ and ${\tilde g}$  being scaling functions and 
$\epsilon=(q-q_c)/f_c$. Typically, $q_c$ is located
by choosing a  quantity that
 intersects  for distinct system
sizes.  For order-disorder phase transitions, the quantity
$U_4$
fulfills the above requirement, 
whose  crossing value $U_0^{*}$  depends on the lattice topology and the 
symmetry properties.
Some papers \cite{tome1,tome2}  have described similar
scaling relation for the entropy production. Close
to the criticality  $\Pi$ and
its first derivative
$\Pi'\equiv d\Pi/dq$ behave    as $\Pi- \Pi_c \sim (q_c-q)^{1-\alpha}$ 
and $\Pi'\sim (q_c-q)^{-\alpha}$, respectively. Above expression states that
$\Pi$ is continuous,  but  the derivative $\Pi'$
diverges  at $q=q_c$.
  Due to finite-size effects, it is reasonable to assume 
that $\Pi'$ behaves as
$\Pi'=N^{\alpha/\nu}{\tilde h}(N^{1/\nu}|\epsilon|)$, with ${\tilde h}$
being an appropriate scaling function.
From the exponents $\beta,\alpha$ and $\gamma$, we wish to check whether 
the hyperscaling relation $\alpha+2\beta+\gamma=2$, fulfilled
in the MFT approach, is also satisfied beyond the MFT.
Here we extend the entropy production analysis for continuous phase
transitions 
in random  complex topologies. 
\section{Applications: The inertial majority vote (MV) model}
\subsection{Model and Definitions}
The previous  predictions will be exemplified
in one of the simplest nonequilibrium phase transition model with steady
states, the majority vote (MV) model
\cite{mario92,chen1},  defined as follows:  Each site $i$ of an arbitrary 
lattice can assume ${\bar q}$ possible integer values 
($\sigma_i=0,1,...,{\bar q}-1$). The dynamics is ruled  by 
the fraction  ${\bar w}_X$ of neighboring nodes
in each one of the ${\bar q}$ states plus 
a local spin dependence $\theta \delta(\sigma_i',\sigma_i)$ (an inertial term),   
${\bar w}_{\sigma_i'}=(1-\theta)\sum_{j=1}^{k}\delta(\sigma_i',\sigma_j)/k+\theta \delta(\sigma_i',\sigma_i)$, with $\sigma_j$ denoting the spin of each one
of the $k$ nearest neighbors of the site $i$.
With probability $1-f$ ($f$ being the misalignment parameter)
the local spin $\sigma_i$ changes to the majority neighborhood spin
$\sigma_i'$ and with complementary probability $f$ the majority rule
is not followed. 
For ${\bar q}=2$ and $\theta=0$, the  MV becomes equivalent to the Ising model in contact
with two heat reservoirs, one being  a source of heat, at infinite temperature,
and  the other a  sink of heat, at zero temperature \cite{mario92}.
The contact with the first occurs with a given probability and with
the second with the complementary probability.
Recent studies  \cite{chen2,pedro,jesus} revealed that
large inertia shifts the phase transition
to a discontinuous one for all values of ${\bar q}$.
An order-disorder phase transition  arises by
increasing $f$, 
whose classification depends on $\theta$
and the lattice connectivity $k$.  
For low ${\bar q}$ (${\bar q}<4$) and $\theta=0$ (inertialess regime), it  is  always continuous \cite{mario92,chen1,chen2},
but the increase of  ${\bar q}$ 
modifies the symmetry properties ($Z_2$ and  $C_{3v}$ for ${\bar q}=2$ and $3$, 
respectively), leading to   different sets of critical exponents.   
The phase transition becomes 
 discontinuous for larger $k$'s when  $\theta$ goes up \cite{chen2,jesus}.
A given $n-$th order parameter moment $\langle m^n \rangle$ is calculated
through the quantity $ \langle m^n \rangle=\langle| \sum_{i=1}^{N}e^{2\pi i \sigma_i/{\bar q}}/N|^n\rangle$, with $\langle...\rangle$ denoting the ensemble average.
The $n=1$ is a
reliable order-parameter since $m >0$ $(=0)$ in the ordered (disordered) phases.
The steady entropy production rate  is  calculated from Eq. (\ref{prod}) through 
the expression
\begin{equation}
  \Pi=\frac{k_B}{N}\left\langle \sum_{j=1}^{{\bar q}-1}\sum_{i=1}^{N}w_i(\sigma) \ln \frac{w_i (\sigma)}{w_i (\sigma^{j})}\right\rangle,
  \label{eqep3}
\end{equation}
with 
$w_i (\sigma)$ and $w_i (\sigma^{j})$ being the transition rate and its reverse,
respectively. The latter is evaluated by taking  transformation
 of $\sigma_i$  to one of its ${\bar q}-1$ distinct values.
For ${\bar q}=2$, the transition
rate above is more conveniently rewritten by taking the
transformation $\sigma_i \rightarrow 2\sigma_i-1$, so that
$w_i(\sigma)$ and $m$  reads
$w_i(\sigma)=\frac{1}{2}[1-(1-2f)\sigma_iS(X)]$ and $m=\langle \sigma_i\rangle$,
 respectively, where
$S(X)$  again denotes the sign function evaluated over the local neighborhood
plus the inertia
$X=(1-\theta)\sum_{j=1}^{k}\sigma_j/k+\theta \sigma_i$. Thus,
in such case $X$ not only depends on the neighborhood, but
also on the local spin $\sigma_i$. The
steady state expression for the absolute $m$  reads
\begin{equation} 
m=(1-2f)\langle {\rm S}(X) \rangle.
\end{equation}

In order to evaluate $\Pi$ from Eq. (\ref{eqep3}) we take the ratio between $w_i (\sigma)$
and its reverse $w_i (\sigma^j)$ given by
\begin{equation}
\frac{w_i (\sigma)}{w_i (\sigma^j)}=
\frac{1-(1-2f)\sigma_i S[\sum_{j=1}^{k}\sigma_j+\frac{k\theta}{1-\theta}\sigma_i]}{1+(1-2f)\sigma_i S[\sum_{j=1}^{k}\sigma_j-\frac{k\theta}{1-\theta}\sigma_i]}.
\label{ep1}
\end{equation}
Inspection of the ratio above reveals that only  local configurations
with $|\sum_{j=1}^{k}\sigma_j|$ greater than $k\theta/(1-\theta)$
will contribute for $\Pi$, since only in these  cases the
ratio is different from 1. Thereby, it can be
rewritten as $w_i (\sigma)/w_i (\sigma^j)=\sigma_iS'(X)\ln [f/(1-f)]$, with $S'(X)$ being the sign function evaluated
only over the subspace of local configurations in which
the ratio is different from $1$ (for  $\theta=0$,
it reduces to the usual sign function). The expression
for $\Pi$ is then given by
\begin{equation}
\Pi=\frac{1}{2}\ln\frac{f}{1-f}\left[\langle \sigma_iS'(X) \rangle-(1-2f)\langle S'^2(X)\rangle\right],
\label{eqw1}
\end{equation}
in such a way that  it depends on
 the averages  $\langle \sigma_iS'(X) \rangle$ and
 $\langle S'^2(X)\rangle$.

\subsection{MFT Results}
The (general) results from Sec. II can be 
straightforwardly applied for the inertialess
regime simply by replacing $q$ and $g(X)$ for $1-2f$ and $S(X)$,
respectively. Although
the main aspects of phase transitions are expected not depending
 on $\theta$,   
in such case it is more convenient to use Eq. (\ref{eqw1}), 
due to the dependence on the local spin. 
The  MFT expression
for $m$ read 
\begin{equation}
m=(1-2f)\left[\langle S\left[X_+\right]\rangle\left(\frac{1+m}{2}\right)-\langle S\left[X_-\right]\rangle\left(\frac{1-m}{2}\right)\right].
\label{eq5}
\end{equation}
 As in Sec. II, for large $k$ the $\langle S\left[X_\pm\right]\rangle$ can be 
calculated from Eq. (\ref{eq4}), but the lower limits $n_{\pm}$
depend on $\theta$ and  are given by
\[n_+=\frac{k(1-2\theta)}{2(1-\theta)}\quad {\rm and} \quad n_-=\frac{k}{2(1-\theta)}.
\] 
Note that both $n_\pm$ reduce to $k/2$ when $\theta=0$.
By performing similar calculations that those
from Sec. II,  Eq. (\ref{eq5}) in the regime of large connectivities becomes
\begin{equation}
m=\frac{(1-2f)\left[{\rm erf(a)}-{\rm erf(b)}\right]}{2-(1-2f)\left[{\rm erf(a)}+{\rm erf(b)}\right]},
\label{eq6}
\end{equation}
where ${\rm erf(x)}$ denotes the error function, with  $a$ and $b$ given by 
\begin{equation}
a=\sqrt {\frac{k}{2}}\left[\frac{\theta}{1-\theta}+m\right] \quad {\rm and} \quad b=\sqrt {\frac{k}{2}}\left[\frac{\theta}{1-\theta}-m\right].
\end{equation}

As performed previously,  the one-site MFT for $\Pi$ is obtained 
 by replacing $\langle \sigma_iS'(X)\rangle $ for
 $\langle \sigma_i\rangle \langle S'(X)\rangle$, so that
\begin{equation}
  \Pi=\frac{1}{2}\ln \frac{f}{1-f} \left[ m\langle S'(X) \rangle-(1-2f)\langle S'^2(X)\rangle\right].
\label{ep21}
\end{equation}

\begin{figure*}
\centering
\includegraphics[scale=0.5]{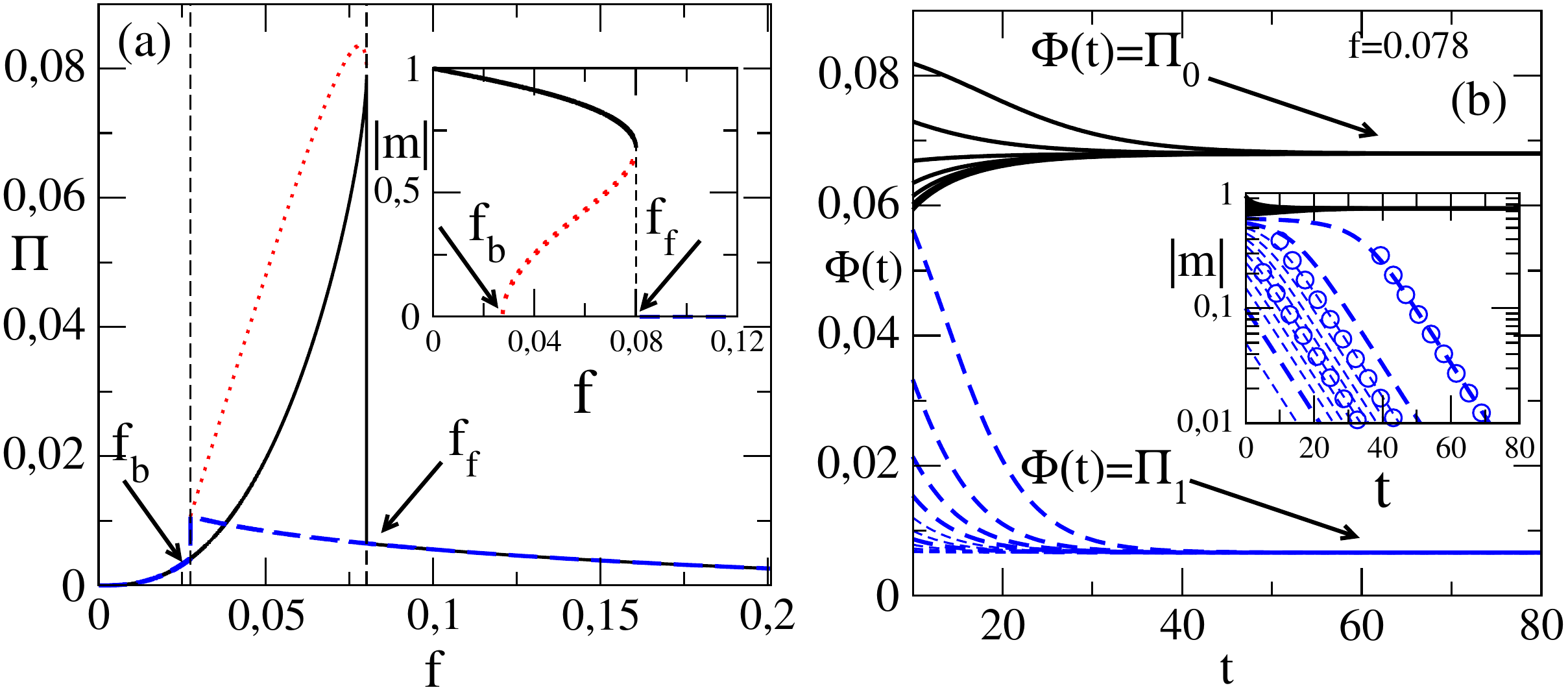}
\caption{Panel $(a)$ depicts the  bistable behavior of $\Pi$
  for $\theta=0.43$ and $k=12$.
Continuous (dashed)  curves denote the stable  solutions for
$m_0>m^{(U)}$ ($m_0<m^{(U)}$).
They coincide 
for $f>f_f$ and $f<f_b$  and
are different for $f_b<f<f_f$. Dotted curves  correspond the unstable solutions
for $f_b<f<f_f$ with $m=m^{(U)}(f)$ if $m_0=m^{(U)}(f)$. Inset: The same but for 
the order-parameter. 
In $(b)$ the time evolution of flux $ \Phi(t)$
for distinct initial configurations and $f=0.078$. Inset:
The time evolution of $m$, where circles correspond to the function $m \sim e^{a(f_b-f)t}$, valid
for $m_0\ll 1$.}
\label{fig2m}
\end{figure*}

  Fig. \ref{fig2m} summarizes the main results for 
the former case for $k=12$ and distinct inertia values. 
As predicted in Sec. II, the order parameter jumps
at $f_f$ and $f_b$ and the discontinuities are also presented
in the entropy production. Along the hysteretic branch, $\Phi(t)$   converges to two well
defined values which are $\Pi_1\equiv\Pi(f,\theta)$ and $\Pi_0\equiv\Pi(m^{(S)},f,\theta)$ in the region $f_b<f<f_f$.  The time evolution
of $m$ follows theoretical prediction $m \sim e^{a(f_b-f)t}$ 
for $m_0\ll 1$ (see inset symbols).

\begin{figure*}
  \centering
  \includegraphics[scale=0.5]{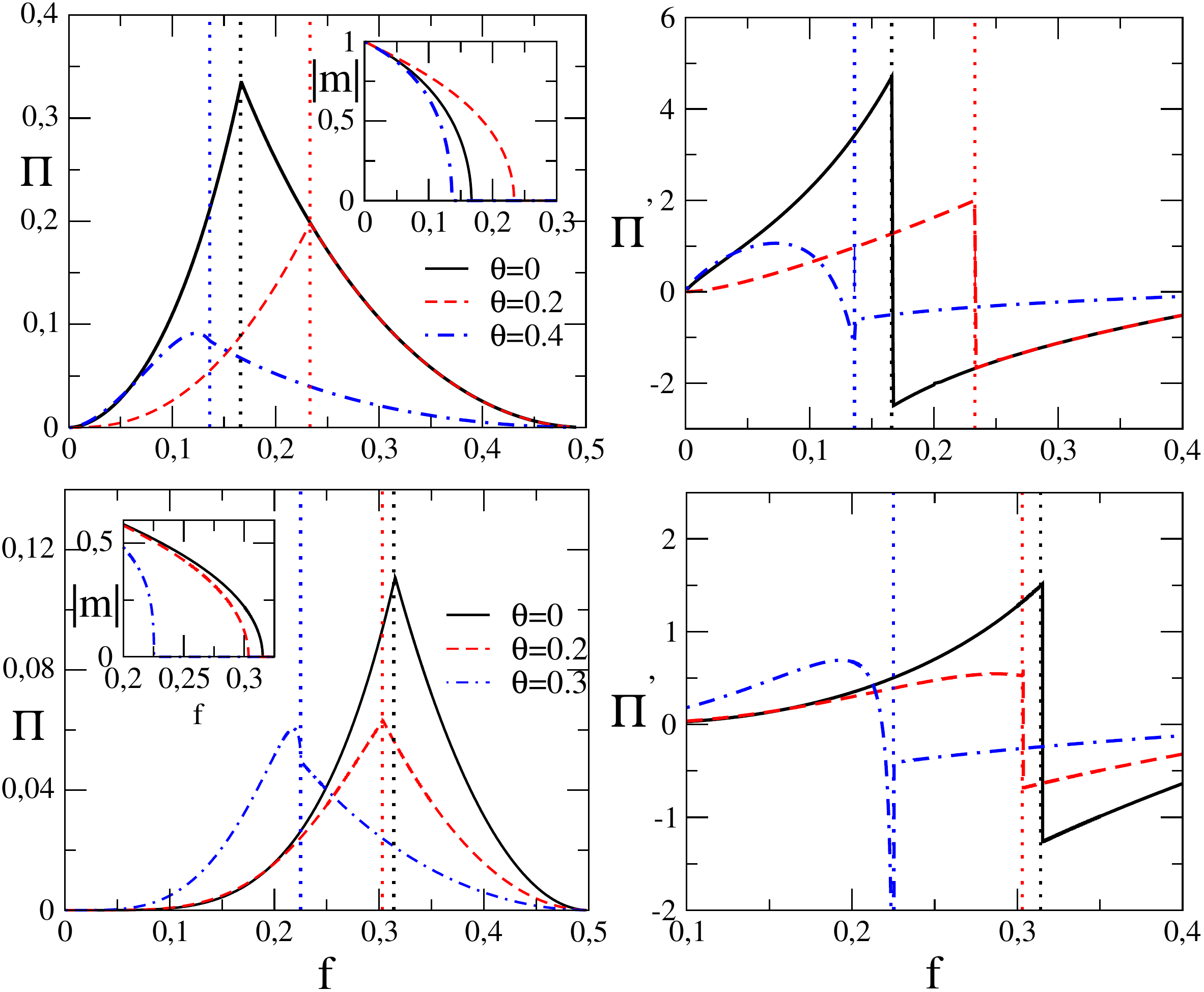}
\caption{Left and right panels: Steady entropy production $\Pi$
  and its derivative $\Pi'$ versus $f$ for low  $\theta$,
   $k=4$ (top) and $k=12$ (bottom), respectively. Inset: The corresponding order
  parameter  versus $f$. Dotted lines denote the associated critical points. }
\label{fig1}
\end{figure*}

Fig. \ref{fig1} exemplifies the main results
for  continuous phase transitions.  In all cases,
the entropy production  increases 
 until a maximum at  $f=f^*$ and then decreases for $f>f^*$. 
 For  the inertialess case or even the low $\theta$, $f^*=f_c$. 
This can be understood by resorting the findings
from Sec. II (for $q=1-2f$ and $g(X)=S(X)$) in which in
 the regime of  large $k$, $m$ and $\Pi$ are given by 
\begin{equation}
m=(1-2f)\mathrm{erf}\Big( m \sqrt{\frac{k}{2}}\Big),
\label{eqm}
\end{equation}
and
\begin{equation}
 \Pi=\frac{1}{2}\ln \frac{f}{1-f} \left[ \frac{m^2}{ 1-2f}-(1-2f) \right],
  \label{eqlk}
\end{equation}
respectively.
At the vicinity of the critical point, where  $m$ 
is expected to be small,  the  right side of Eq. (\ref{eqm})  
can be expanded in Taylor series, allowing us
to rewrite $m$ solely in terms of $f$ and $k$:
\begin{equation}
 m\sim {\sqrt \frac{12}{k}}(f_c-f)^{1/2},
\label{opc}
\end{equation}
where $\beta_{mf}=1/2$   is the  critical exponent
and
\begin{equation}
f_c=\frac{1}{2}\left\{1-\sqrt {\frac{\pi}{2k}} \right\},
\end{equation}
is the critical point.
 From Eq.  (\ref{opc}), 
$\Pi$ behaves  as
$\Pi \approx \frac{1-2f}{2}\ln \frac{1-f}{f} \left[1-\frac{12}{k}\frac{f_c-f}{(1-2f)^2} \right]$ and
   $\Pi=\frac{1-2f}{2}\ln \frac{1-f}{f}$ for 
$f \rightarrow f_c^-$ and  $f>f_c$, respectively, and
hence $\Pi$ is continuous at the criticality. 
 Despite this, its first derivative
  $\Pi'$ jumps from 
$\frac{1}{2}{\sqrt\frac{\pi}{2k}}\ln\frac{1+{\sqrt\frac{\pi}{2k}}}{1-{\sqrt\frac{\pi}{2k}}}$ to   $\frac{12}{\pi}\ln\frac{1+{\sqrt\frac{\pi}{2k}}}{1-{\sqrt\frac{\pi}{2k}}}$, hence consistent with the exponent $\alpha_{mf}=0$. 
By increasing $\theta$ (see e.g 
$\theta=0.4$ and $0.3$ for $k=4$ and $k=12$,  respectively), 
the maximum of $\Pi$ does not coincide with the jump of $\Pi'$
nor the order-parameter vanishing. 
Thereby the present results (together with the general description in Sec. II)
 unifies the description in the MFT context,
in which the criticality is not necessarily marked
by a peak in the entropy production but related to  a peculiar
behavior of its first derivative.

\begin{figure}[h]
\centering
\includegraphics[scale=0.32]{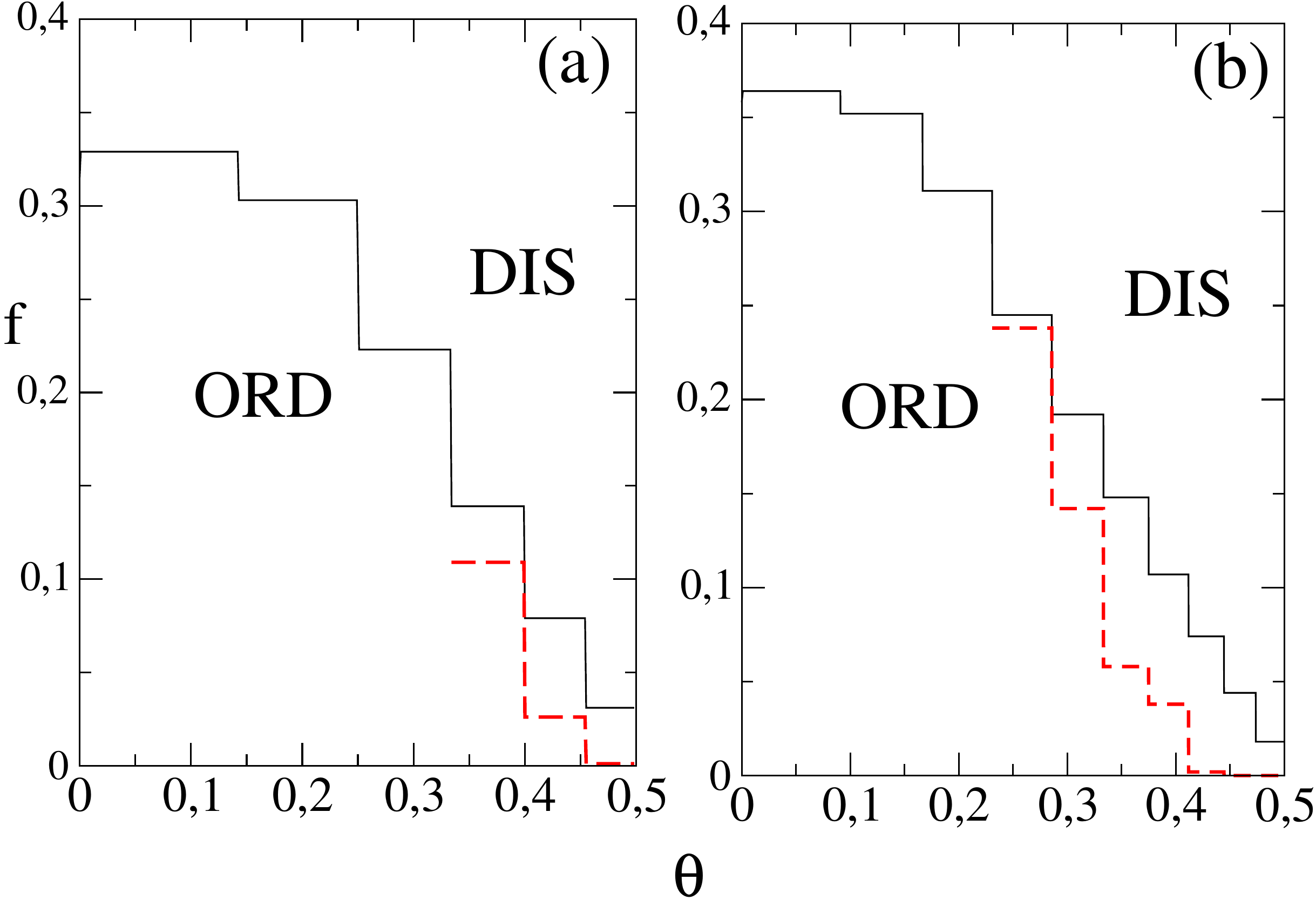}
\caption{Panels $(a)$ and $(b)$ show the mean-field
  phase diagrams for $k=12$ and $k=20$ through analysis of
  entropy production. ORD (DIS) denote the ordered (disordered) phases, whereas
  continuous and dashed lines, correspond to the values of
  $f_f$ and $f_b$, respectively. They coincide  for continuous
transitions, but are different for discontinuous ones.}
\label{mftp}
 \end{figure}

Lastly, in Fig. \ref{mftp} we plot the phase diagrams for $k=12$ and $k=20$
evaluated through the distinct entropy production signatures.
We see that both phase transition location
 and its classification are in full agreement with
those obtained from order-parameter analysis (see e.g. Fig. 1 in 
Refs. \cite{chen2}).

A final comment concerns that the limit  $k \rightarrow \infty$
corresponds to the 
complete graph regime. In this case, the expression
for  $m$ and $\Pi$ become 
\begin{equation} 
m=\frac{(1-2f)\left[ S(\frac{\theta}{1-\theta}+m)- S(\frac{\theta}{1-\theta}-m)\right]}{ 2-(1-2f)\left[ S(\frac{\theta}{1-\theta}+m)+ S(\frac{\theta}{1-\theta}-m)\right]},
\label{aq}
\end{equation}
and
\begin{equation}
\Pi=\ln\frac{f}{1-f}\left\{m-(1-2f)Y_p\right\}, 
\end{equation}
respectively, where $Y_p=\{(1+m)S[m+\theta/(1-\theta)]-(1-m)S[m-\theta/(1-\theta)]  \}/2$. By combining the above relation with Eq. (\ref{aq}), it follows that
$\Pi=0$ and thus there is no entropy production in the complete graph
case. The reversible character of the inertialess MV in the
complete graph has  already been
presented in Ref. \cite{fron} and  our analysis not only confirms it but
also extends  for the inertial regime. 

\subsection{Beyond the MFT: Numerical Results in regular
and complex structures}
Numerical simulations will be performed for
distinct lattices structures and neighborhoods. 
All studied structures are quenched, i.e., 
they do not change during the simulation of the model. 
For a given network topology with  $N$, $f$, and $\theta$ held fixed, a site $i$ is randomly chosen,
and its spin value $\sigma_i$ is updated 
($\sigma_i \rightarrow \sigma_i'$) according to ${\bar w}_{\sigma_i'}=(1-\theta)\sum_{j=1}^{k}\delta(\sigma_i',\sigma_j)/k+\theta \delta(\sigma_i',\sigma_i)$, with $\sigma_j$ denoting the spin of each one
of the $k$ nearest neighbors of the site $i$. With probability $1-f$,
 $\sigma_i$ changes to the majority neighborhood spin
$\sigma_i'$ and with complementary probability $f$ the majority rule
is not followed. 
A Monte Carlo (MC) step corresponds to $N$ updating spin trials. 
After repeating the above dynamics a sufficient
number of MC steps (in order of $10^{6}$ MC steps), the system attains 
a nonequilibrium steady state.

Random regular networks have been generated through a configuration model scheme
  \cite{boll} described as follows: For a system with 
$N$ nodes and connectivity $k$, we first
 start with a set of $Nk$ points, distributed in $N$ groups,
 in which each one contains exactly $k$ points. Next, one chooses
 a random pairing of the points between groups and then
creates a network linking the nodes $i$ and $j$
if there is a pair containing points in the $i$-th and $j$-th sets until
$Nk/2$ pairs (links) are obtained. If
 the resulting network configuration present a loop or
duplicate links, the above process  is restarted.

\begin{figure}[h]
\centering
\includegraphics[scale=0.3]{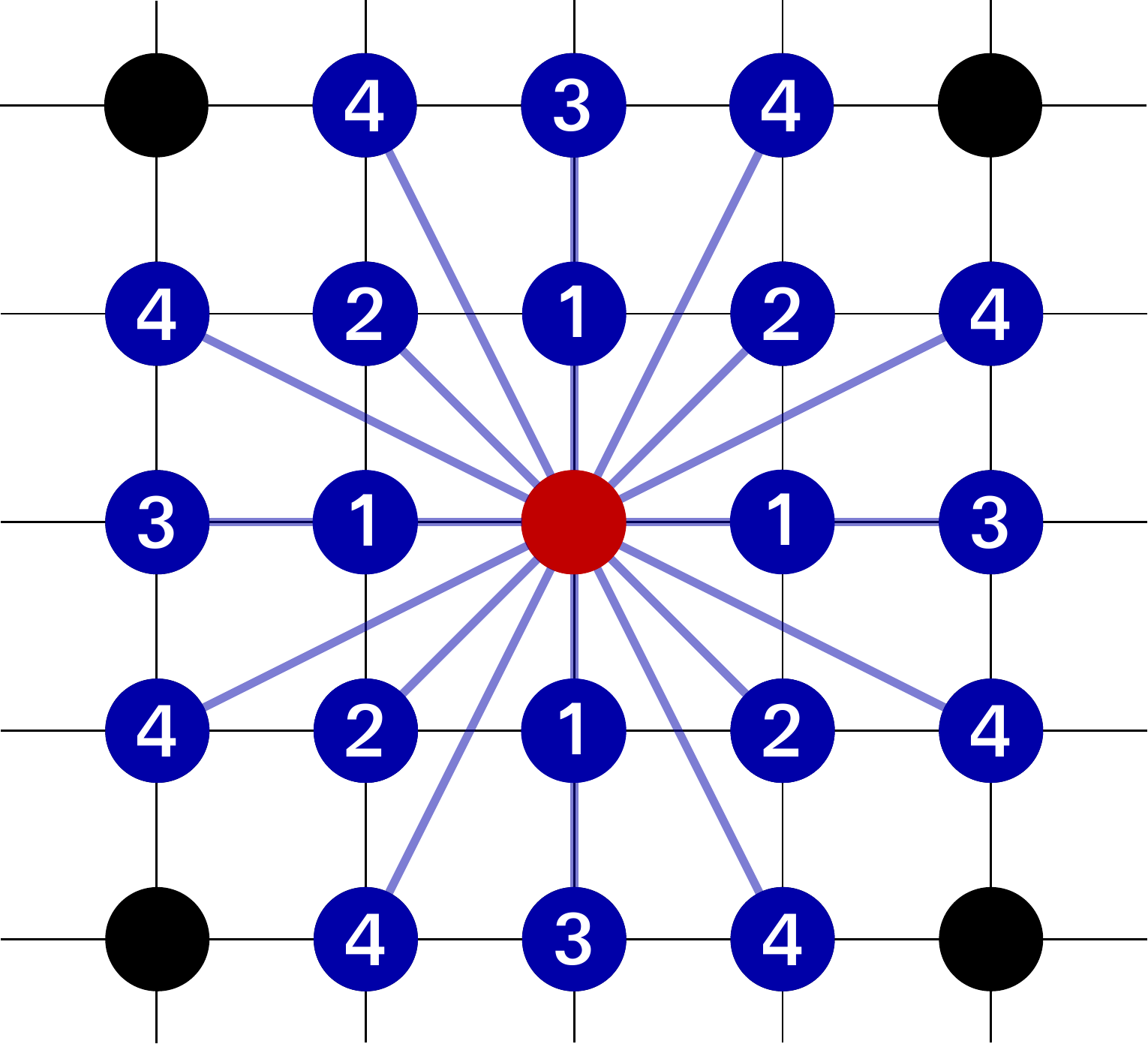}
\caption{Local configuration for a bidimensional
  lattice  with central site (red) and its
  first  $(1)$, second $(2)$,  third $(3)$ and  fourth $(4)$ next neighbors.}
\label{figs2}
 \end{figure}

The increase of connectivity $k$ in bidimensional topologies
is accomplished by extending the range of interaction
neighborhood. For example, $k=4,8,12$ and $20$ includes interaction
between the  first, first and second, first  to third and first to
fourth next neighbors, respectively, as sketched in Fig. \ref{figs2}.

\subsubsection{Discontinuous phase transitions}

\begin{figure*}
\centering
\includegraphics[scale=0.4]{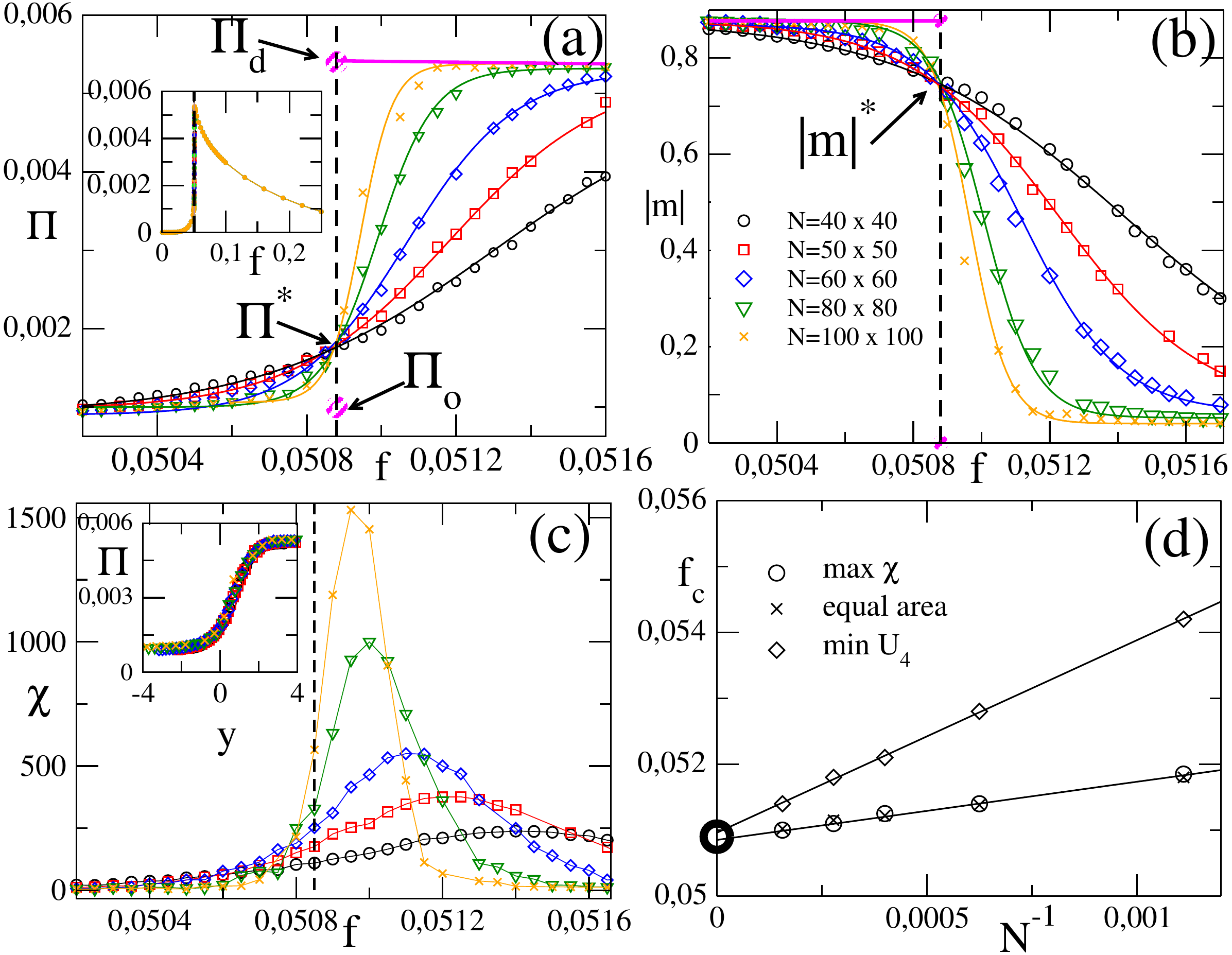}
\caption{Bidimensional lattice with $k=20$ and $\theta=0.375$.
  Panels $(a)$-$(c)$ show the steady  $\Pi$,  the order parameter
  $|m|$ 
  and the variance $\chi$ versus $f$, respectively, for
  distinct system sizes at the vicinity
  of phase coexistence. Dashed lines: Crossing point among entropy production curves.
  Continuous lines in $(a)$ and $(b)$ correspond to
  the theoretical description, Eq. (\ref{opr}). Top and bottom insets: $\Pi$ for larger sets
  of $f$ and collapse of data by taking the relation
$y=(f-f_0)N$, respectively.
  In $(d)$, the  plot of the maximum of $\chi$, minimum of $U_4$
and equal area order-parameter probability distribution
 versus $N^{-1}$.   }
\label{fig2-1}
\end{figure*}

\begin{figure*}
\centering
\includegraphics[scale=0.45]{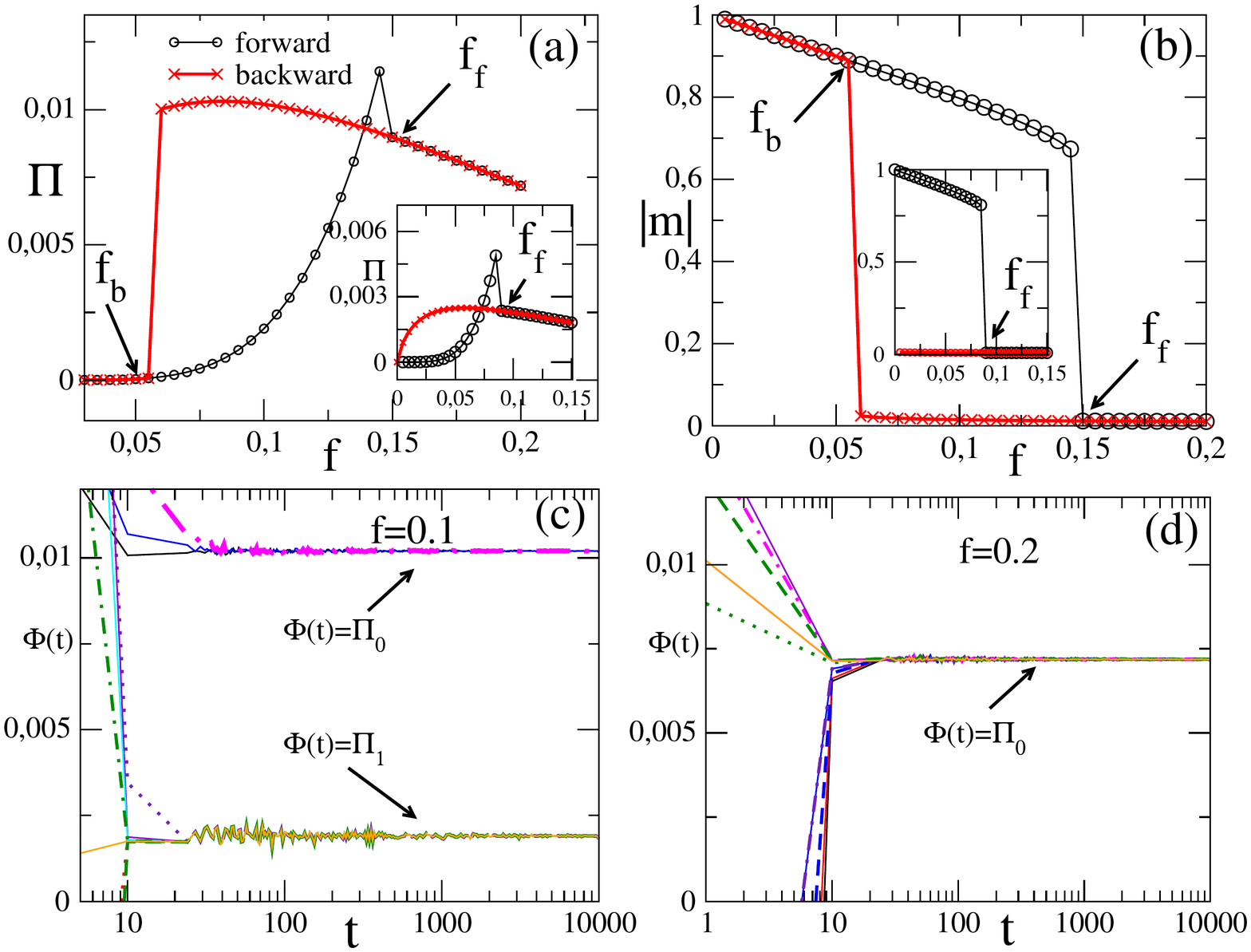}
\caption{Panels $(a)$ and $(b)$ 
show the steady  $\Pi$ and $|m| $ versus $f$ for
$k=20$, $\theta=0.3$ for the random-regular (RR) case with $N=10^4$.
Black and red curves correspond to the forward and backward ``trajectories'',
respectively.
Inset:  The same but
for $\theta=0.375$. In $(c)$ and $(d)$, 
the time evolution of $\Phi(t)$ for distinct initial conditions
$m_0$ for $f_b<f=0.10<f_f$  and $f=0.20>f_f$, respectively.
 For larger inertia  values (inset),
the bistability extends over $0\le f\le f_f$, also viewed from the behavior
of steady $\Pi$.}
\label{fig2}
\end{figure*}

Fig. \ref{fig2-1} 
exemplifies such predictions  for the MV in bidimensional lattices
with $k=20$ and $\theta=0.375$.
The entropy production curves follow the
theoretical predictions (continuous lines in panels $(a)$ and $(b)$) from Eqs. (\ref{opr}) and (\ref{crossing}), whose
intersection  among curves (panels $(a)$ and $(b)$)  occurs at 
$f_0=0.05084(5)$, in excellent agreement
with estimates obtained from standard techniques \cite{fsize2},
$0.0509(1)$ (maximum of $\chi$), $0.0510(1)$ (minimum of
$U_4=1-\langle m^{4} \rangle/3\langle m^{2} \rangle^{2}$) and $0.0509(1)$ (equal area order-parameter distribution $P_N(m)$)-see e.g. panel $(d)$.
Collapse of all data by taking the transformation $y=(f-f_0)N$ (inset)
reinforces the reliability of Eq. (\ref{opr}) for describing
 $\Pi$ at the phase coexistence region.
Out of the scaling regime   ($f>f_0$ for large $N$),
$\Pi$ depends solely on the control parameters ($f$ and $\theta$ for the
MV), as can be seen
in the upper inset of Fig. \ref{fig2-1}.
The crossing in both order
parameter and entropy production not only discerns  
the behavior from regular and complex topologies (see e.g. Fig. \ref{fig2})
but also discontinuous
and continuous phase transitions (see e.g. Fig. \ref{fig2c}).

\begin{figure}
\centering
\includegraphics[scale=0.35]{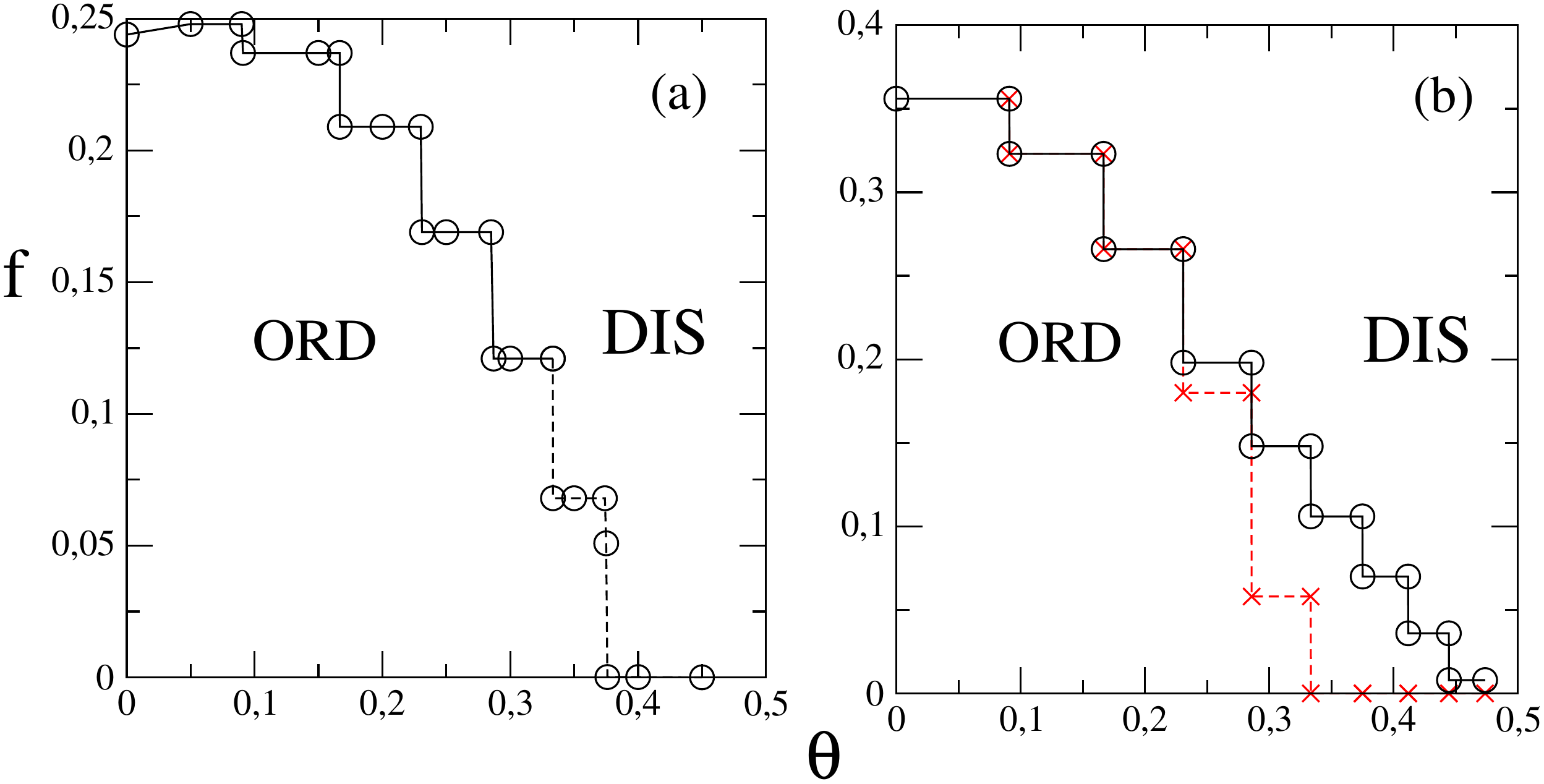}
\caption{Panels  $(a)$ and $(b)$ show the 
  phase diagrams for $k=20$ for regular and RR structures,
  respectively, through analysis of
  entropy production. ORD (DIS) denote the ordered (disordered) phases
  and continuous (dashed) lines correspond to continuous (discontinuous)
  phase transitions. In $(b)$,  circles ($\times$) 
  correspond to the increase (decrease) of $f$ starting from
  an ordered (disordered) phase.}
\label{fig4cc}
 \end{figure}

Conversely,  Fig. \ref{fig2} depicts the main results for the MV in
  a random-regular (RR)  topology, 
for $k=20, \theta=0.3$ and $N=10^4$.   In such case, the
entropy production  reveals typical
signatures from aforementioned complex networks: the existence of a
hysteretic loop [panel $(a)$]  located at
  the interval $f_b=0.055<f<f_f=0.15$, in full
  equivalence with the order-parameter 
branch [panel $(b)$] \cite{chen2,jesus}.  
The phase diagrams, calculated from the entropy production analysis, are shown
in Fig. \ref{fig4cc} for both regular and complex networks.

\begin{figure*}
\centering
\epsfig{file=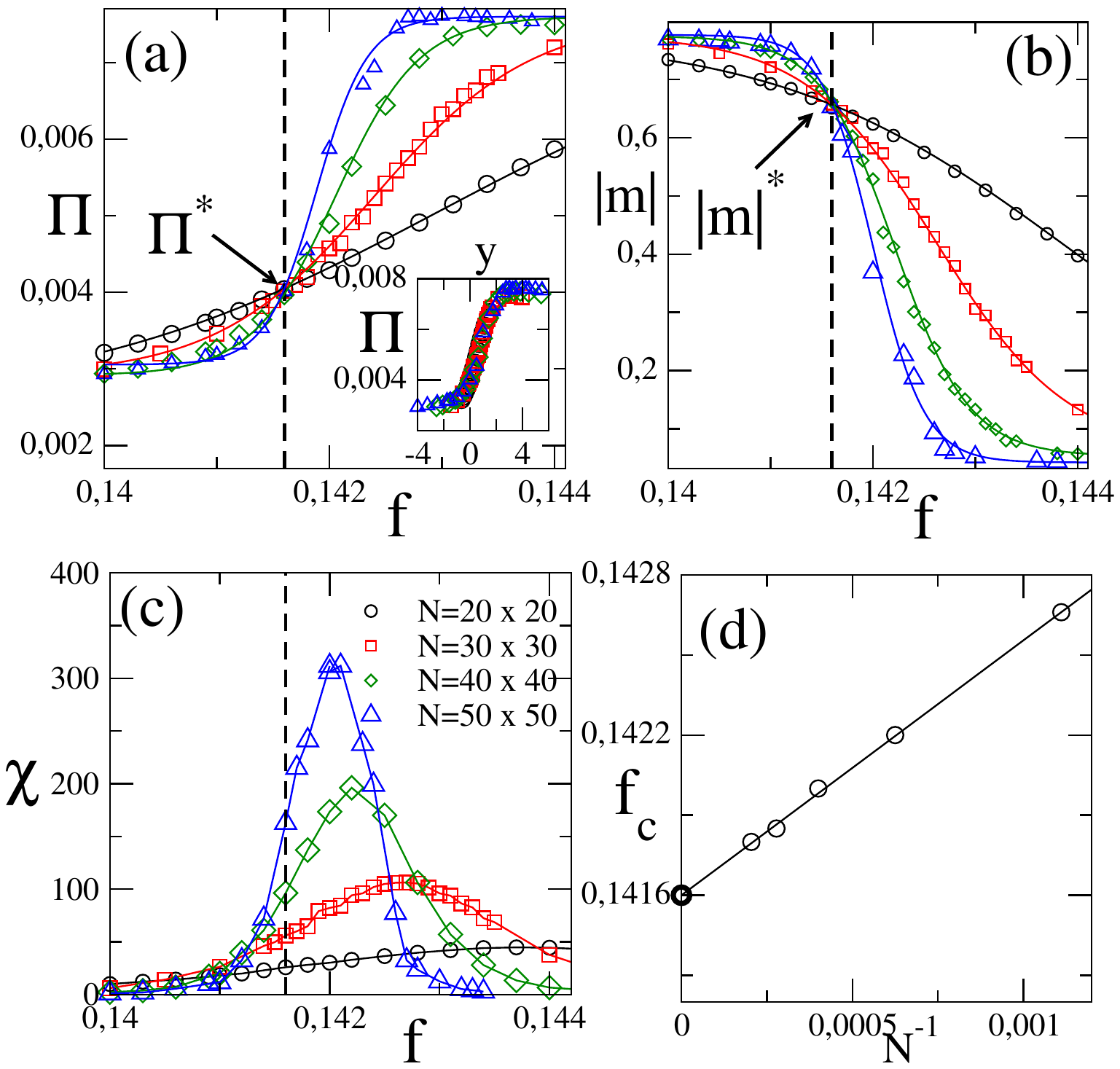,width=14cm,height=18cm}
\caption{Regular lattice for $k=20$
  and $\theta=0.32$:
  Panels $(a)$-$(c)$ depict the steady  $\Pi$,  the order parameter
  $|m|$   and the variance $\chi$ versus $f$, respectively, for
  distinct system sizes at the vicinity
  of phase coexistence. Dashed lines: Crossing point among entropy production
  curves. Continuous lines in $(a)$ and $(b)$ are
    the theoretical description presented in Eq. (\ref{opr}). 
  Inset: Collapse of data by taking
  the relation $y=(f-f_0)N$.   In $(d)$, the  plot of
  the maximum of $\chi$ versus $N^{-1}$.}
\label{fig5c}
 \end{figure*}
\begin{figure*}
\centering
\includegraphics[scale=0.45]{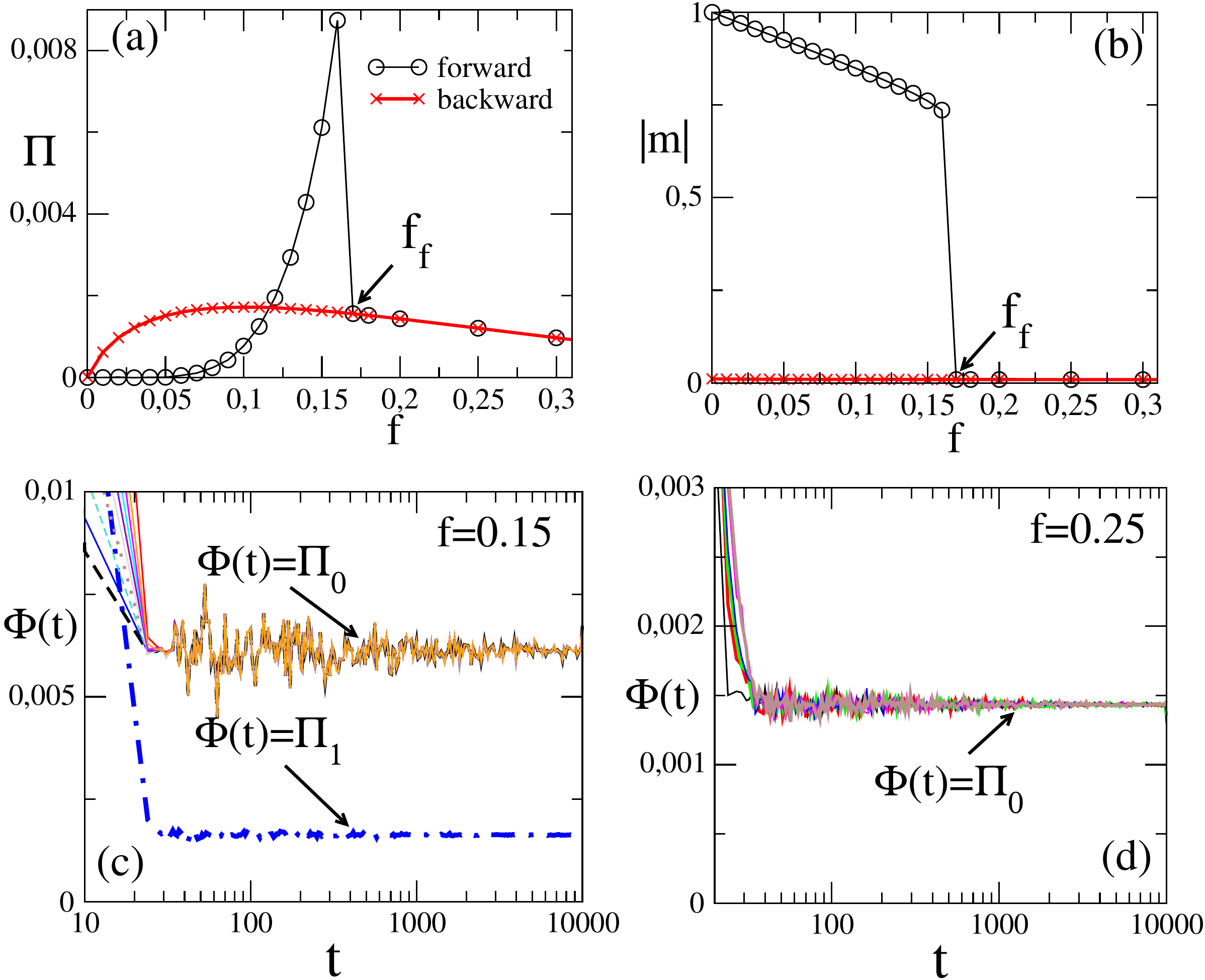}
\caption{For the RR structure, panels $(a)$ and $(b)$ 
show the steady  $\Pi$ and $|m|$ versus $f$ for
  $k=20$ and $\theta=0.35$.  In $(c)$ and $(d)$, 
the time evolution of $\Phi(t)$ for distinct initial conditions
for $f=0.15$ (bistable loop) and $f=0.25$ (disordered phase), respectively. }
\label{fig3d}
 \end{figure*}

Figs.  \ref{fig5c} and \ref{fig3d} depict the main results
for the bidimensional  and random-regular structures for ${\bar q}=3$,
in which the $C_{3v}$ symmetry leads to an entirely different critical behavior
from  the ${\bar q}=2$ case. However, the phase coexistence portraits  are
analogous to the previous ones, including the existence of
bistability (complex networks), crossing among
curves at the transition point ($f_0=0.14160(5)$) and scaling with
the system volume (regular structures), thereby reinforcing
the robustness of our findings  at discontinuous phase transitions.

\subsubsection{Continuous phase transitions}
\begin{figure*}
\centering
\includegraphics[scale=0.45]{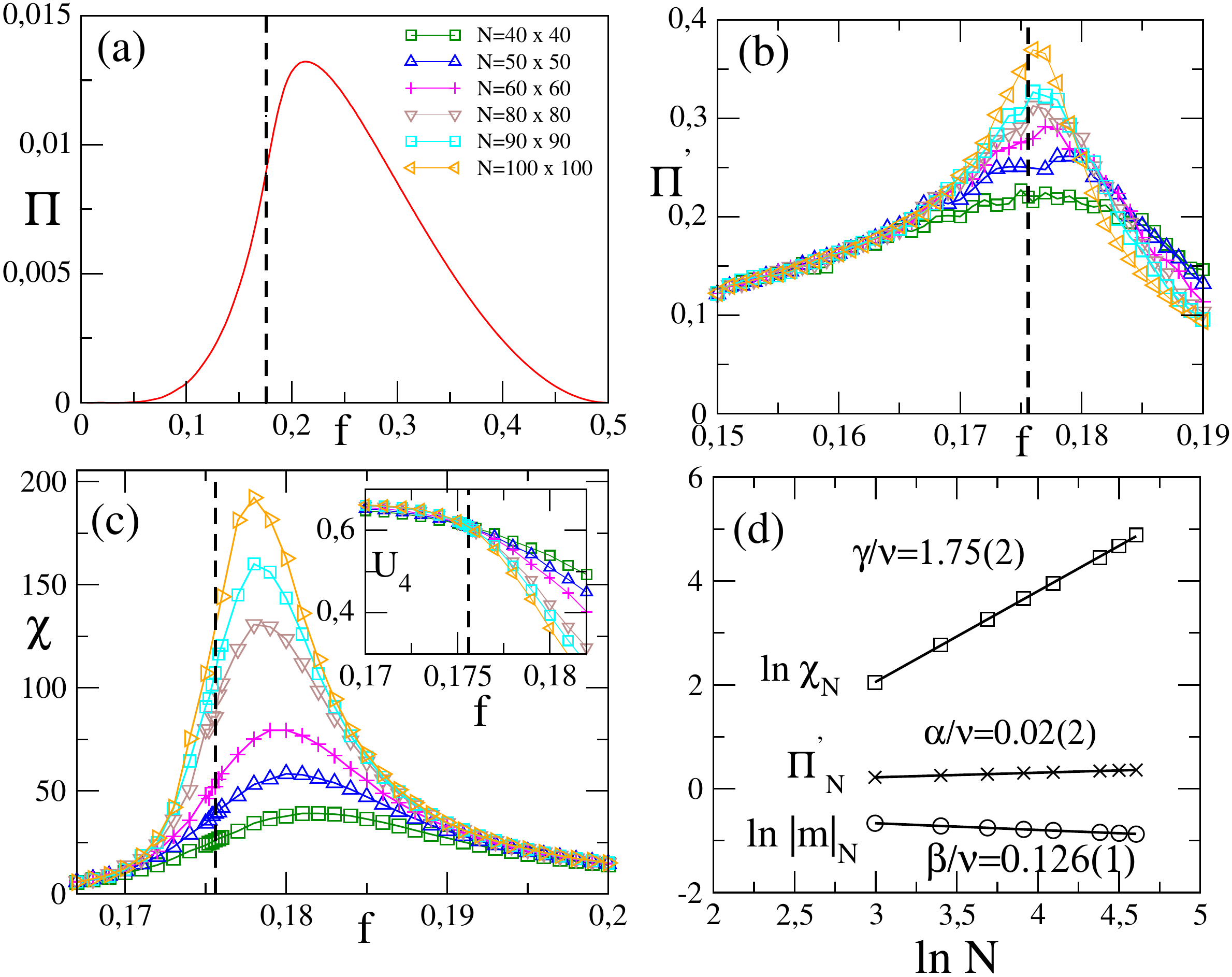}
\caption{Regular lattice for interactions between the first to
  the third next neighbors ($k=12$)
  and $\theta=0.2$: Panels $(a)$, $(b)$ and $(c)$ depict the 
  entropy production $\Pi$, its derivative $\Pi'$ the variance $\chi$ versus $f$,
  respectively for distinct system sizes. Inset: The same but for 
 fourth-order reduced cumulant $U_4$. Dashed lines denote the critical point
$f_c$ evaluated through the crossing
among $U_4$ curves. In $(d)$, the $\ln \chi_{N}$, $\ln |m|_{N}$
  and  $\Pi_N'$ versus $\ln N$ at $f=f_c$. }
\label{fig3c}
\end{figure*}

  Previous results  show that irrespectively the
value of $\theta$  \cite{jesus},
the phase transition remains continuous in regular structures when $k<20$
whose critical  exponents are consistent with the values
 $\beta=1/8$, $\gamma=7/4$ and $1/\nu=1$ \cite{mario92}.
Fig. \ref{fig3c}  illustrates    continuous
phase transition traits 
in terms of the entropy production.
Although   $\Pi$ is finite 
in the critical point [panel $(a)$],
$\Pi'$ increases without limits as $N\rightarrow \infty$  [panel $(b)$].
For finite systems, $\Pi'_N$ evaluated at $f=f_c$ increases with  $\ln N$, 
consistent to a logarithmic divergence in which one
associates the exponent  $\alpha=0$ [panel $(d)$]. 

\begin{figure*}
  \centering
\includegraphics[scale=0.43]{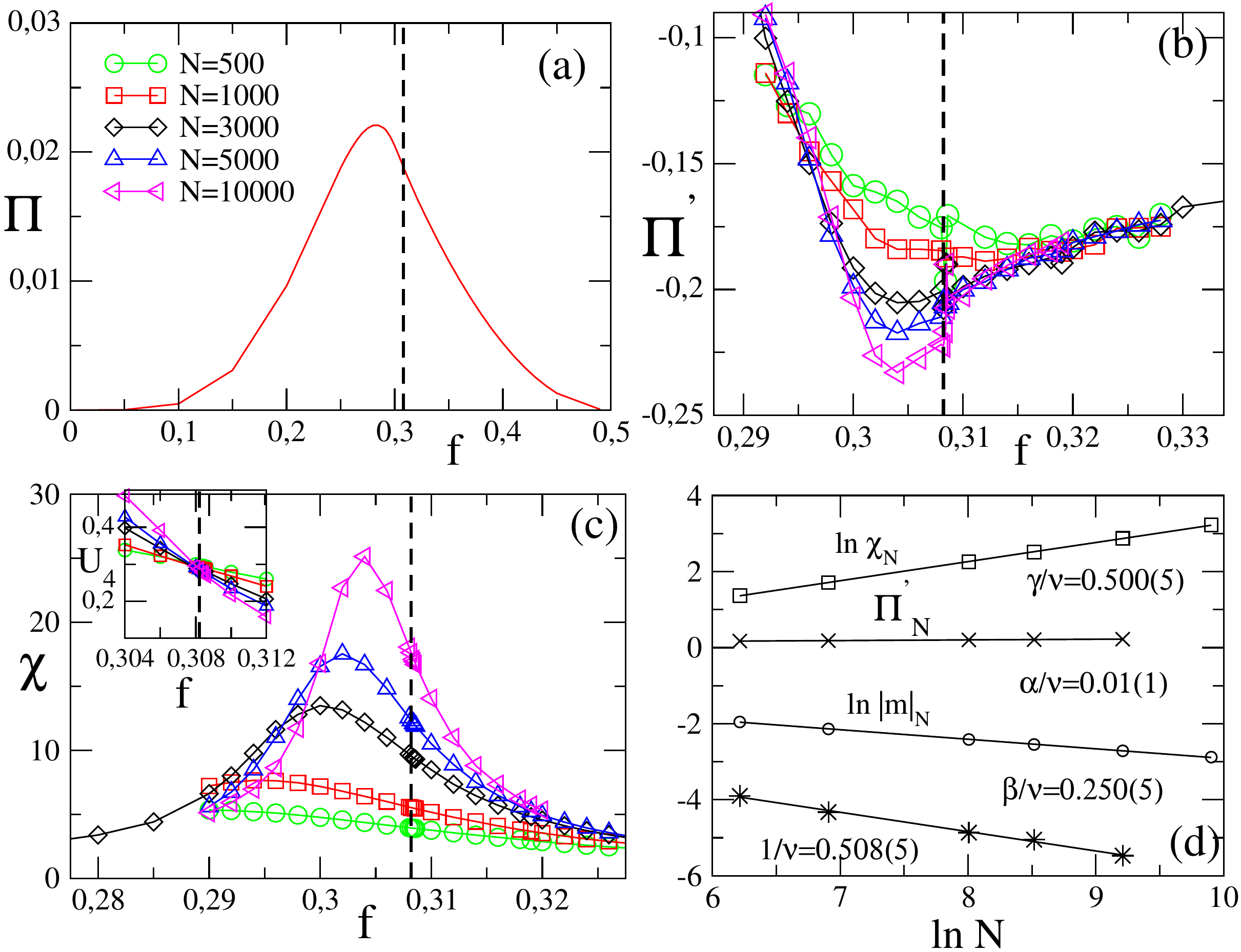}
\caption{Random regular (RR) network with $k=12$ and $\theta=0$:
  Panels $(a)-(c)$ show the
  steady entropy production $\Pi$,  its derivative
  $\Pi'$ and $\chi$ versus $f$ for distinct system sizes respectively. 
  Inset: the reduced cumulant $U_4$ vs $f$.
  Dashed lines denote the critical point $f_c$ evaluated through the crossing
among $U_4$ curves.
Panel $(d)$ depicts the plot of $\ln \chi_{N}$, $\ln |m|_{N}$, $\Pi'_{N}$ and
$\ln (f_c-f_N)$ versus $\ln N$ at $f=f_c$. }
\label{fig2c}
\end{figure*} 
Fig. \ref{fig2c} extends the analysis  for RR structures.
In that case, the critical behavior
follows the exponents 
$\beta/\nu=1/4$, $\gamma/\nu=1/2$ and $1/\nu=1/2$ \cite{pereira}, rather
different from  $\beta=1/2$, $\gamma=1$ and $1/\nu=2$ (MFT) and those
from regular lattices (Fig. \ref{fig3c}).
In similarity to the bidimensional
case, $\Pi(f_c)$ is finite and  $\Pi'_N(f_c)$ increases with  $\ln N$,
which is also
consistent to a logarithmic divergence and with the exponent 
$\alpha=0$. As in Sec. II, 
such conclusions are reinforced by appealing
to the hyperscaling relation $\alpha+2\beta+\gamma=2$. Having the
values of $\beta$ and $\gamma$, we reobtain  in both cases $\alpha=0$.
 Lastly, the  ${\bar q}=3$  case is 
characterized in regular lattices by the  critical exponents $\beta=1/9$ and $\gamma=13/9$.
According to the hyperscaling relation,   the
exponent associated with the entropy production should read $\alpha=1/3$.
Very recently, the value  $\alpha=0.32(2)$ has been confirmed 
from numerical simulations in Ref. \cite{tomeo}, in full accordance with our 
theoretical predictions.
The present analysis not only puts on firmer basis the behavior
of entropy production at the criticality
but also extends the hyperscaling relation for 
nonequilibrium phase transitions.

\section{ Conclusions}
Based on  general considerations,
the
description of entropy production traits for continuous and
discontinuous (practically  unexplored) phase transitions was presented. 
Our main findings are that continuous and discontinuous phase transitions
can be classified through  specific (well defined) entropy production traits
in the realm of MFT and beyond  MFT.
Our approach embraces fundamental aspects comprising
the influence of the lattice topology and symmetry properties.
At the phase coexistence, the entropy production presents a
discontinuity in a single (and well defined) point 
in regular lattices, whereas a hysteretic loop is portrayed in
complex networks. The former case is also characterized by
  the existence of a crossing point among  entropy production curves for distinct system
  sizes.
A general description of entropy production in the framework of  mean-field
theory for systems with $Z_2$ symmetry was presented.
Our work is a relevant step in trying to  unify the description of
 nonequilibrium phase transitions  through
 a key indicator of system irreversibility.
As a final comment, it would be interesting 
to consider the critical behavior of entropy production (and its allied quantities)
for systems displaying other symmetries and universality classes, 
in order to verify the reliability
of finite size ideas presented here.


\section{Acknowledgment}
C. E. F. and P. E. H. acknowledge the financial support from FAPESP
under grants No 2018/02405-1 and 2017/24567-0, respectively.


\begin{thebibliography}{99a}
 \bibitem{prigo} I. Prigogine,
   {\it Introduction to Thermodynamics of Irreversible
     Processes}, 2nd ed. (Wiley, New York, 1961).
   \bibitem{groot} S. R. de Groot and P. Mazur, {\it Non-Equilibrium
     Thermodynamics} (North-Holland, Amsterdam, 1962).
     \bibitem{mariobook} T. Tom\'e and M. J. de Oliveira,
{\it Stochastic Dynamics and Irreversibility} (Springer, Cham, 2015).
\bibitem{seinf} U. Seifert, Rep.  Prog.  Phys. {\bf 75}, 126001
(2012).
\bibitem{tome1} L. Crochik and T. Tom\'e, Phys. Rev. E {\bf 72}, 057103 (2005).
\bibitem{tome2}  T. Tom\'e and M. J. de Oliveira,  Phys. Rev. Lett.
   {\bf 108}, 020601 (2012).
\bibitem{tome3} T. Tom\'e and M. J. de Oliveira,  Phys. Rev. E.
  {\bf 91}, 042140 (2015).
\bibitem{barato} Y. Zhang and A. C. Barato, J. Stat. Mech.   {\bf 2016},
  113207 (2016).
  
\bibitem{andrae} B. Andrae, J. Cremer, T. Reichenbach and E. Frey,
  Phys. Rev. Lett. {\bf 104}, 218102 (2010).

\bibitem{gaspard} P. Gaspard, J. Chem. Phys. {\bf 120}, 8898 (2004).
\bibitem{qian} Hao Ge  and Hong Qian, J. R. Soc. Int.  {\bf 8}, 107 (2011).
\bibitem{imparato} A. Imparato, New J. Phys. {\bf 17},  1025004 (2015).
  \bibitem{shim} P. S. Shim, H. M. Chun and J. D. Noh, Phys. Rev E {\bf 93}, 012113 (2016).
\bibitem{esposito1} T. Herpich, J. Thingna and M. Esposito, Phys. Rev. X {\bf 8}, 031056 (2018).
\bibitem{esposito2} T. Herpich and M.  Esposito, Phys. Rev. E {\bf 99}, 022135 (2019).

  
  
  \bibitem{mandal}D. Mandal,  K. Klymko and M. R. DeWeese, 
Phys. Rev. Lett {\bf 119}, 258001 (2017).
\bibitem{landi} M Brunelli, L. Fusco, W. Wieczorek, J. Hoelscher-Obermaier,
  G. T. Landi, F. L. Semião, A. Ferraro, N. Kiesel, T. Donner, G. De Chiara and M. Paternostro, Phys. Rev. Lett {\bf 121}, 16064 (2018).
  \bibitem{schn}  J. Schnakenberg, Rev. Mod. Phys. {\bf 48}, 571 (1976).
\bibitem{wetting}  A. C. Barato and H. Hinrichsen, J. Phys. A {\bf 45},
  115005 (2012).

\bibitem{mario92} M. J. de Oliveira, J. Stat. Phys. {\bf 66}, 273
(1992).

\bibitem{chen1} H. Chen, C. Shen, G. He, H. Zhang and Z. Hou, Phys,
Rev. E {\bf 91}, 022816 (2015).
\bibitem{chen2} H. Chen, C. Shen, H. Zhang, G. Li, Z. Hou
  and J. Kurths, Phys Rev. E {\bf 95}, 042304 (2017).
\bibitem{pedro} 
  P. E. Harunari, M. M. de Oliveira, and C. E. Fiore,
  Phys Rev. E {\bf 96}, 042305 (2017).
  \bibitem{jesus} 
  J. M. Encinas, P. E. Harunari, M. M. de Oliveira and C. E. Fiore,
Sci. Rep. {\bf 8}, 9338 (2018).
\bibitem{paula} See e.g. P. V. Mart\'in, J. A. Bonachela, S. A. Levin, 
and M. A. Mu\~noz,
Proc. Natl. Acad. Sci. USA {\bf 112}, E1828 (2015).

\bibitem{fsize} M. M. de Oliveira, M. G. E. da Luz, and C. E. Fiore,
Phys. Rev. E {\bf 92}, 062126 (2015).

 \bibitem{fsize2} M. M. de Oliveira, M. G. E. da Luz and C. E. Fiore,
  Phys. Rev. E {\bf 97}, 060101(R) (2018).

\bibitem{challa}  M. S. S. Challa, D. P. Landau, and K. Binder,
  Phys. Rev. B {\bf 34}, 1841 (1986).
 
 \bibitem{haye} R. Ziener, A. Maritan and H. Hinrichsen, J. Stat. Mech.
   2015, P08014 (2015).

   \bibitem{pereira}  L. F. C. Pereira and F. G. B. Moreira, 
      Phys. Rev. E {\bf 71}, 016123 (2005).
   \bibitem{boll}  B. Bollob\'as,  {\it Europ. J. Combinatorics},
     {\bf 1}, 311 (1980).
\bibitem{stama} M. Pineda and M. Stamatakis, Entropy {\bf 20}, 811 (2018).
\bibitem{prl2011} See e.g. J. G\'omez-Garde\~nes, S. G\'omez, A. Arenas and Y.
Moreno, Phys. Rev. Lett. {\bf 106}, 128701 (2011).
\bibitem{fron} A. Fronczak and P. Fronczak,  Phys. Rev. E {\bf 96},
  012304 (2017).
\bibitem{tomeo} O. A. Barbosa and T. Tom\'e, to be published in J.Phys. A (2019).
 \end{thebibliography}
\end{document}